\documentclass[12pt]{amsart}
\usepackage{color,cite}
\newcommand\R{{\mathbb{R}}}

\pdfoutput=1
\usepackage[pdftex]{graphicx}
\usepackage{tikz}
\usepackage{subcaption}
\begin{document}
\DeclareGraphicsExtensions{.pdf,.png,.mps}

\newtheorem{theorem}{Theorem}[section]
\newtheorem{lemma}[theorem]{Lemma}
\newtheorem{Pro}[theorem]{Proposition}
\newtheorem{definition}[theorem]{Definition}
\newtheorem{proposition}[theorem]{Proposition}
\newtheorem{remark}[theorem]{Remark}
\newtheorem{corollary}[theorem]{Corollary}
%\newtheorem{Conj}[Theorem]{Conjecture}
%\newtheorem{Prob}[Theorem]{Problem}
%\newtheorem{Ques}[Theorem]{Question}

%documentclass[12pt,thmsa]{article}
%\usepackage{amsfonts}

%%%%%%%%%%%%%%%%%%%%%%%%%%%%%%%%%%%%%%%%%%%%%%%%%%%%%%%%%%%%%%%%%%%%%%%%%%%%%%%%%%%%%%%%%%%%%%%%%%%
%\usepackage{sw20jart}
\title[Open books]{Spectra of ``fattened'' open book structures}
\author{James E. Corbin}
\address{Department of Mathematics, Texas A\& M University, College Station, TX}
\email{jxc8004@math.tamu.edu}
\author{Peter Kuchment}
\address{Department of Mathematics, Texas A\& M University, College Station, TX}
\email{kuchment@math.tamu.edu}
\thanks{The work of both authors has been partially supported by the NSF DMS-1517938 grant.}
\dedicatory{Dedicated to the memory of great mathematician and friend Victor Lomonosov}
\date{\today}
\maketitle
%\tableofcontents
\begin{abstract}
We establish convergence of spectra of Neumann Laplacian in a thin neighborhood of a branching 2D  structure in 3D to the spectrum of an appropriately defined operator on the structure itself. This operator is a 2D analog of the well known by now quantum graphs. As in the latter case, such considerations are triggered by various physics and engineering applications.
\end{abstract}

%%%%%%%%%%%%%%%%%%%%%%%%%%%
\section*{Introduction}
\begin{figure}[ht!]
 \begin{center}
\begin{tikzpicture} [scale = 2.2]
\draw [ultra thick] (-1, -1) -- (0,0);
\draw [thick] plot [smooth] coordinates { (-1, -1) (-1.05, -0.35)   (-2.2, 0.4) };
\draw [thick] plot [smooth] coordinates { (-1, -1) (-1.3, -0.4)  (-1.7, -0.35) (-2.3, -0.5) };
\draw [thick] plot [smooth] coordinates {(-1,-1)  ( -1.5, -0.92) (-2.4, -0.9) };
\draw [thick] plot  [smooth] coordinates {( -1,-1) (-0.5, -0.7) (0.5, -0.6)};
\draw [shift ={(1,1)}] [thick] plot [smooth] coordinates { (-1, -1) (-1.05, -0.35)   (-2.2, 0.4)  };
\draw [shift ={(1,1)}] [thick] plot [smooth] coordinates { (-1, -1) (-1.3, -0.4)  (-1.7, -0.35) (-2.3, -0.5) };
\draw [shift ={(1,1)}] [thick] plot [smooth] coordinates {(-1,-1)  ( -1.5, -0.92) (-2.4, -0.9) };
\draw [shift ={(1,1)}] [thick] plot  [smooth] coordinates {( -1,-1) (-0.5, -0.7) (0.5, -0.6)};
\draw [thick] (-2.2, 0.4)--(-2.2 +1, 0.4 + 1);
\draw [thick] (-2.3, -0.5) -- (-2.3 +1,  -0.5+1);
\draw [thick] (-2.4, -0.9) -- (-2.4+1, -0.9+1);
\draw [thick]  (0.5, -0.6)--  (0.5+1, -0.6+1);
\node at (0.5, -0.1) {$M_4$};
\node at (-1.12, 1) {$M_1$};
\node at (-0.8, 0.25) {$M_2$};
\node at (-1.7, -0.75) {$M_3$};
\
\end{tikzpicture}
\end{center}
  \caption{An open book structure with ``\textbf{pages}'' $M_k$ meeting at a ``\textbf{binding}.''}\label{F:OBS}
\end{figure}
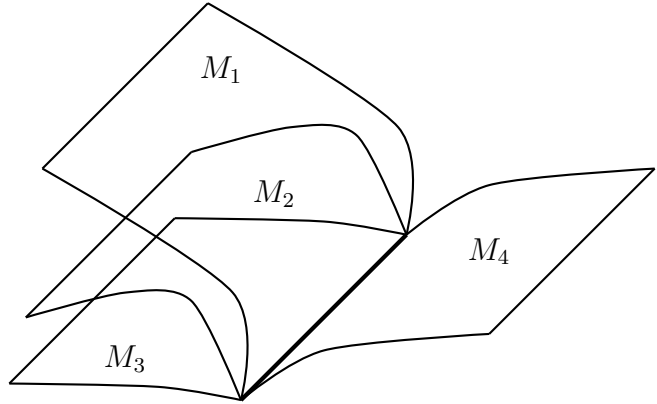

\begin{figure}[ht!]
\begin{center}
\begin{tikzpicture} [scale = 1.5]
\shade [ball color =gray!40, opacity = 0.3] (-1,0) circle (2);
\shade [ball color = gray!40, opacity = 0.3] (1,0) circle(1.5);
\draw (0.4,-1.37) arc (270:450:-0.4 and 1.37);
\draw [dashed] (0.4,-1.37) arc (270:450:0.4 and 1.37);
\node at (-2, 0) {$M_1$};
\node at (-0.25,0) {$M_2$};
\node at (2,0) {$M_4$};
\node at (1, 1 ) {$M_3$};
\node at (0.2,0) {$E$};
\end{tikzpicture}
\end{center}
\caption{A transversal intersection of two spheres yields an open book structure with
four pages and a circular binding. The requirement of absence of zero-dimensional strata prohibits
adding a third sphere with a generic triple intersection. Tangential contacts of spheres are also disallowed.}\label{F:spheres}
\end{figure}
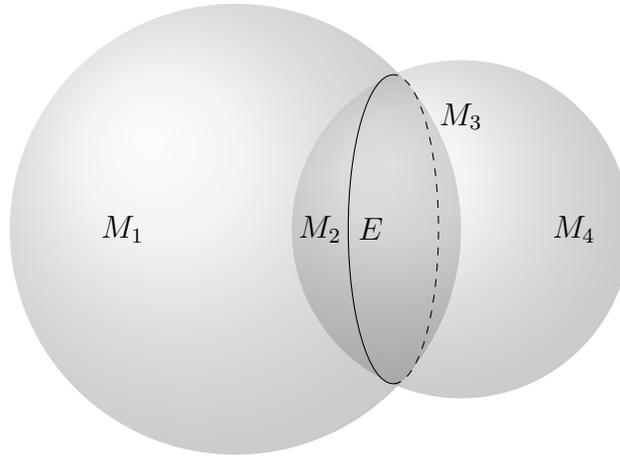

%%%%%%%%%%%%%%%%%%%%%%%%%%%
We consider a compact sub-variety $M$ of $\R^3$ that locally (in a neighborhood of any point) looks like either a smooth submanifold or an ``open book'' with smooth two-dimensional ``pages'' meeting transversely along a common smooth one-dimensional ``binding,''\footnote{We do not provide here the general definition of what is called Whitney stratification, see e.g. \cite{Arnold,Lu,strat,Whit}, resorting to a simple description through local models.} see Fig. \ref{F:OBS}.
Clearly, any compact smooth submanifold of $\R^3$ (with or without a boundary) qualifies as an open book structure with a single page. Another example of such structure is shown in Fig. \ref{F:spheres}.

%The $2D$-strata need neither be contractible, nor orientable.

A ``fattened'' version $M_\epsilon$ of $M$ is
an (appropriately defined) $\epsilon$--neighborhood of $M$, which we call a ``fattened open book structure.''

Consider now the Laplace operator $-\Delta$ on the domain $M_\epsilon$ with Neumann boundary conditions (``\textbf{Neumann Laplacian}''), which we denote $A_\epsilon$.
As a (non-negative) elliptic operator on a compact manifold, it has discrete finite multiplicity spectrum $\lambda^\epsilon_n:=\lambda_n(A^\epsilon)$ with the only accumulation point at infinity.
The result formulated in this work is that when $\epsilon\to 0$, each eigenvalue $\lambda_n^\epsilon$ converges to the corresponding eigenvalue $\lambda_n$ of an operator $A$ on $M$, which acts as $-\Delta_M$ (2D Laplace-Beltrami) on each 2D stratum (\textbf{page}) of $M$, with appropriate junction conditions along 1D strata (\textbf{bindings}).

Similar results have been obtained previously for the case of fattened graphs (see \cite{KZ,RS2}, as well as books \cite{BerKuc,Post} and references therein), i.e. $M$ being one-dimensional.

The case of a \emph{smooth} submanifold $M\subset \R^3$ is not that hard and has been studied well under a variety of constraints set near $M$ (e.g., \cite{FH,Grieser,BerKuc,WRM}). Having singularities along strata of lower dimensions significantly complicates considerations, even in the quantum graph case \cite{Grieser,DellAn06,DellAn07,DellAn15,KuISAAC,KZ,KZ2,WRM,RS2,S,EH}.

Our considerations are driven by the similar types of applications (see, e.g. \cite{BerKuc,AnGr,ES,FK1,FK3,FK4,FK5,Fr,FW,FWopb,KuPBG,KuPBG,WRM,RS1,RS2,RS3,RS4,RueS}), as in the graph situation.

The Section \ref{S:notions} contains the descriptions of the main objects: open book structures and their fattened versions, the Neumann Laplacian $A$, etc. The next Section \ref{S:result} contains formulation of the result. The proof is reduced to constructing two families of ``averaging'' and ``extension'' operators. This construction is even more technical than in the quantum graph case and will be provided in another, much longer text. The last Section \ref{S:remarks} contains the final remarks and discussions.

In this article the results are obtained under the following restrictions: the width of the fattened domain shrinks ``with the same speed'' around all strata; no ``corners'' (0D strata) are present; the pages intersect transversely at the bindings. Some of them will be removed in a further work.

%%%%%%%%%%%%%%%%%%%%%%%%
\section{The main notions}\label{S:notions}
%%%%%%%%%%%%%%%%%%%
%Here we introduce the main geometric objects and differential operators to be studied.
%%%%%%%%%%%%%%%%%%%%%%%%%%%%%%%%
\subsection{Open book structures}\label{SS:OBS}
%%%%%%%%%%%%%%%%%%%%%%%

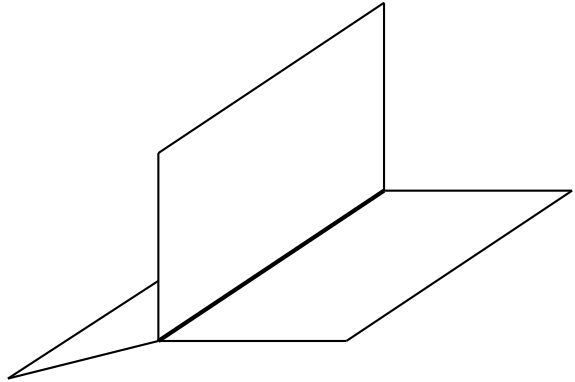
\begin{figure}[ht!]
  \centering
  \begin{tikzpicture}
\draw [ultra thick] (0,0) -- (3,2);
\draw [thick] (0,0) -- (0,2.5);
\draw [thick] (3,2) -- (3,4.5);
\draw [thick] (0,2.5) -- (3,4.5);
\draw [thick] (2.5,0) -- (5.5,2);
\draw[ thick] (3,2) -- (5.5,2);
\draw [thick] (0,0) -- (2.5,0);
\draw [thick]  (-2, -0.5) -- (0,0);
\draw [thick] (-2, -0.5) -- (0, 0.8);
\end{tikzpicture}
\caption{A local model of a binding neighborhood}
\label{F:locmod}
\end{figure}

Simply put, an open book structure\footnote{One can find open book structures in a somewhat more general setting being discussed in algebraic topology literature, e.g. in \cite{Ranic,Winkel}.} $M$ is connected and consists of finitely many connected, compact smooth submanifolds (with or without boundary) of $\R^3$ (\textbf{strata}) of dimensions two and one, such that they only intersect along their boundaries and each stratum's boundary is the union of some lower dimensional strata \cite{strat}.
We also assume that the strata intersect at their boundaries transversely.
In other words, locally $M$ looks either as a smooth surface, or an ``open book'' with pages meeting at a non-zero angle at a ``binding.''
Up to a diffeomorphism, a neighborhood of the binding looks like in Fig. \ref{F:locmod}.

%%%%%%%%%%%%%%%%%%%%%%%%%%%%
\subsection{The fattened structure}\label{SS:fatten}
%%%%%%%%%%%%%%%%%%%%%%

We can now define the \textbf{fattened open book structure} $M_\epsilon$. %If $M$ were a smooth surface in $\R^3$, we could ``fatten'' it by considering its $\epsilon$-neighborhood $M_\epsilon$.

Let us remark first of all that there exists $\epsilon_0>0$ so small that for any two points $x_1,x_2$ on the same page of $M$, the closed intervals of radius $\epsilon_0$ normal to $M$ at these points do not intersect.
This ensures that the $\epsilon<\epsilon_0$-fattened neighborhoods do not form a ``connecting bridge'' between two points that are otherwise far away from each other along $M$. We will assume that in all our considerations $\epsilon < \epsilon_0$, which is not a restriction, since we will be interested in the limit $\epsilon\to 0$.

We denote the ball of radius $r$ about $x$ as $B(x,r)$.
%In the planned sequel to this article we consider a more general variable width neighborhood, with a smooth function $w(x)$ on $M$ controlling the variable width\footnote{When $w=1$, this boils down to the standard neighborhoods $M_\epsilon$.}.
%
%The \textbf{fattened domain} $M_\epsilon$ for some $\epsilon>0$ consists of all points at the distance of order $\epsilon$ from $M$, plus possibly ``fatter'' neighborhoods $E_{m,\epsilon}$ of the bindings $E_m$. Let us make this more precise.
%
%
%\begin{remark} \label{L:epsilon0}
%
%\end{remark}

\begin{definition} \label{D:Me}
Let $M$ denote an open book structure in $\R^3$ and $\epsilon_0>0$, as defined above.
We define for any $\epsilon<\epsilon_0$ the corresponding fattened domain $M_\epsilon$ as follows:
\begin{equation}
M_\epsilon := \bigcup_{x \in M} B(x, \epsilon)
\end{equation}
%The similar notation $R_\epsilon$ will be used for the fattened version of any subset $R\in\R^3$.
\end{definition}

\subsection{Quadratic forms and operators}\label{SS:qforms}
%%%%%%%%%%%%%%%%%%%%%

We adopt the standard notation for Sobolev spaces (see, e.g. \cite{Ma}). Thus, $H^1(\Omega)$ denotes the space of square integrable with respect to the Lebesgue measure functions on a domain $\Omega \subset \mathbb{R}^n$ with square integrable first order weak derivatives.
\begin{definition} \label{D:Qe}
Let $Q_\epsilon$ be the closed non-negative quadratic form with domain $H^1(M_\epsilon)$, given by
\begin{equation}
 Q_{\epsilon} (u ) = \int_{M_{\epsilon}} | \nabla u |^2 \, dM_\epsilon
\end{equation}
We also refer to $Q_{\epsilon}(u)$ as the \textbf{energy} of $u$.
\end{definition}

This form is associated with a unique self-adjoint operator $A_\epsilon$ in $L_2(M_\epsilon)$. The following statement is standard (see, e.g. \cite{EE,Ma}):

\begin{Pro} \label{P:QeAe}
The form $Q_\epsilon$ corresponds to the \textbf{Neumann Laplacian} $A_\epsilon = -\Delta$ on $M_\epsilon$ with its domain consisting of functions in  $H^2(M_\epsilon)$ whose normal derivatives at the boundary $\partial M_\epsilon$ vanish.

Its spectrum $\sigma(A_\epsilon)$ is discrete and non-negative.
\end{Pro}

Moving now to the limit structure $M$, we equip it with the surface measure $dM$ induced from $\R^3$.

\begin{definition} \label{D:Q}
Let $Q$ be the closed, non-negative quadratic form (\textbf{energy}) on $L_2(M)$ given by
\begin{equation}
Q (u ) = \sum_k \int_{M_k} | \nabla_{M_k} u |^2 \, dM
\end{equation}
with domain $\mathcal{G}^1$ consisting of functions $u$ for whose $Q(u)$ is finite and that are continuous across the bindings between pages $M_k$ and $M_{k'}$:
\begin{equation}\label{E:cont}
u|_{\partial M_k \cap E_m} = u|_{\partial M_{k'} \cap E_m}.
\end{equation}
Here $\nabla_{M_k}$ is the gradient along $M_k$ and restrictions in (\ref{E:cont}) to the binding $E_m$ coincide as elements of $H^{1/2}(E_m)$.
\end{definition}
Unlike the fattened graph case, by the Sobolev embedding theorem \cite{EE} the restriction to the binding is not continuous as an operator from $\mathcal{G}^1$ to $C(E_m)$, it only maps to $H^{1/2}(E_m)$. This distinction significantly complicates the analysis of fattened stratified surfaces in comparison with fattened graphs.

\begin{Pro} \label{Ae}
The operator $A$ associated with the quadratic form $Q$ acts on each $M_k$ as
\begin{equation}
A u :=
-\Delta_{M_k} u,
\end{equation}
with the domain $\mathcal{G}^2$ consisting of functions on $M$ such that
the following conditions are satisfied:
\begin{itemize}
\item
\begin{equation}
||u||_{L_2(M)}^2 +  ||A  u||_{L_2(M)}^2  < \infty,
\end{equation}
continuity across common bindings $E_m$ of pairs of pages $M_k,M_{k'}$:
\begin{equation}
u|_{\partial M_k \cap E_m} = u|_{\partial M_{k'} \cap E_m},
\end{equation}
\item \textbf{Kirchhoff condition} at the bindings:
\begin{equation} \sum_{k: \partial M_k \supset E_m  }  D_{\nu_k} u (E_m) = 0,
\end{equation}
\end{itemize}
where $-\Delta_{M_k}$ is the Laplace-Beltrami operator on $M_k$ and $D_{\nu_k}$ denotes the normal derivative to $\partial M_k$ along $M_k$.

The spectrum of $A$ is discrete and non-negative.
\end{Pro}

The proof is simple, standard, and similar to the graph case. We thus omit it.

%%%%%%%%%%%%%%%%%%%%%%
\section{The main result}\label{S:result}
%%%%%%%%%%%%%%%%%%%%%%%

\begin{definition}
We denote the ordered in non-decreasing order eigenvalues of $A$ as $\{\lambda_n\}_{n\in \mathbb{N}}$, and those of $A_\epsilon$ as $\{\lambda_n^\epsilon \}_{n \in \mathbb{N}}$.

For a real number $\Lambda$ not in the spectrum of $A_\epsilon$, we denote by $\mathcal{P}_{\Lambda}^\epsilon$ the spectral projector of $A_{\epsilon}$ in $L_2(M_\epsilon)$ onto the spectral subspace corresponding to the half-line $\{\lambda\in\R\, |\, \lambda < \Lambda\}$.

Similarly, $\mathcal{P}_{\Lambda}$ denotes the analogous spectral projector for $A$.
We then denote the corresponding (finite dimensional) spectral subspaces as $\mathcal{P}_{\Lambda}^\epsilon L_2(M_\epsilon)$ and $\mathcal{P}_{\Lambda} L_2(M)$  for $M_\epsilon$ and $M$ respectively.
\end{definition}%

We now introduce two families of operators needed for the proof of the main result.

\begin{definition} \label{D:Je}
A family of linear operators $J_{\epsilon}$
from $H^1(M_{\epsilon})$ to $\mathcal{G}^1$ is called \textbf{averaging operators} if for any $\Lambda \notin \sigma(A_\epsilon)$ there is an $\epsilon_0$ such that for all $\epsilon \in (0, \epsilon_0]$ the following conditions are satisfied:
\begin{itemize}
\item For $u \in \mathcal{P}_{\Lambda}^\epsilon L_2(M_{\epsilon})$, $J_\epsilon$ is ``nearly an isometry'' from $L_2(M_\epsilon)$ to $L_2(M)$ with an $o(1)$ error, i.e.
\begin{equation}\label{E:J1}
\bigg| \, \big| \big| u \big| \big|_{L_2(M_{\epsilon})}^2 - \big| \big| J_\epsilon u \big| \big|_{L_2(M)}^2 \, \bigg| \leq o(1) \big| \big| u \big| \big|_{H^1(M_{\epsilon})}^2
\end{equation}
where $o(1)$ is uniform with respect to $u$.
\item For $u \in \mathcal{P}_{\Lambda}^\epsilon L_2(M_{\epsilon})$, $J_\epsilon$ asymptotically ``does not increase the energy,'' i.e.
\begin{equation}\label{E:J2}
 Q( J_\epsilon u) - Q_{\epsilon} ( u )  \leq o(1) Q_{\epsilon}(u)
\end{equation}
where $o(1)$ is uniform with respect to $u$.
\end{itemize}
\end{definition}

\begin{definition} \label{D:Ke}
A family of linear operators $K_{\epsilon}$ from  $\mathcal{G}^1$ to $H^1(M_{\epsilon})$ is called \textbf{extension operators} if for any $\Lambda \notin \sigma(A)$ there is an $\epsilon_0$ such that for all $\epsilon \in (0, \epsilon_0]$ the following conditions are satisfied:
\begin{itemize}
\item  For $u \in  \mathcal{P}_{\Lambda} L_2(M)$, $K_\epsilon$ is ``nearly an isometry'' from $L_2(M)$ to $L_2(M_\epsilon)$ with $o(1)$ error, i.e.
\begin{equation}\label{E:K1}
\bigg| \, \big| \big| u \big| \big|_{L_2(M)}^2 - \big| \big| K_\epsilon u \big| \big|_{L_2(M_{\epsilon})}^2 \, \bigg| \leq o(1) \big| \big| u \big| \big|_{\mathcal{G}^1}^2
\end{equation}
where $o(1)$ is uniform with respect to $u$.
\item For $u \in \mathcal{P}_{\Lambda} L_2(M) $, $K_\epsilon$ asymptotically ``does not increase'' the energy, i.e.
\begin{equation}\label{E:K2}
Q_{\epsilon}( K_\epsilon u)- Q ( u )   \leq o(1) Q(u)
\end{equation}
where $o(1)$ is uniform with respect to $u$.
\end{itemize}
\end{definition}

Existence of such averaging and extension operators is known to be sufficient for spectral convergence of $A_\epsilon$ to $A$ (see \cite{Post}). For the sake of completeness, we formulate and prove this in our situation.
\begin{theorem} \label{T:spec}
Let $M$ be an open book structure and its fattened partner $\{M_\epsilon\}_{\epsilon \in (0, \epsilon_0]}$ as defined before.
Let $A$ and $A_\epsilon$ be the operators on $M$ and $M_\epsilon$ as in Definitions \ref{P:QeAe} and \ref{Ae}.

Suppose there exist averaging operators $\{J_\epsilon\}_{\epsilon \in (0, \epsilon_0]}$ and extension operators $\{K_\epsilon\}_{\epsilon \in (0, \epsilon_0]}$ as stated in Definitions \ref{D:Je} and \ref{D:Ke}.

Then, for any $n$ $$\lambda_n(A_\epsilon) \mathop{\rightarrow}_{\epsilon \to 0} \lambda_n(A).$$
\end{theorem}

We start with the following standard (see, e.g. \cite{ReSi}) min-max characterization of the spectrum.
\begin{proposition}\label{P:Rayleigh}
Let $B$ be a self-adjoint non-negative operator with discrete spectrum of finite multiplicity and $\lambda_n(B)$ be its eigenvalues listed in non-decreasing order. Let also $q$ be its quadratic form with the domain $D$. Then
\begin{equation}\label{E:minmax}
\lambda_n(B)=\mathop{min}\limits_{W\subset D}\,\,\mathop{max}\limits_{x\in W\setminus \{0\}}\frac{q(x,x)}{(x,x)},
\end{equation}
where the minimum is taken over all $n$-dimensional subspaces $W$ in the quadratic form domain $D$
\end{proposition}

\textbf{Proof of Theorem \ref{T:spec}} now employs Proposition \ref{P:Rayleigh} and the averaging and extension operators $J,\,K$ to ``replant'' the test spaces $W$ in  (\ref{E:minmax}) between the domains of the quadratic forms $Q$ and $Q_\epsilon$.

Let us first notice that due to the definition of these operators (the near-isometry property), for any fixed finite-dimensional space $W$ in the corresponding quadratic form domain, for sufficiently small $\epsilon$ the operators are injective on $W$. Since we are only interested in the limit $\epsilon\to 0$, we will assume below that $\epsilon$ is sufficiently small for these operators to preserve the dimension of $W$. Thus, taking also into account the inequalities (\ref{E:J1})-(\ref{E:K2}), one concludes that on any fixed finite dimensional subspace $W$ one has the following estimates of Rayleigh ratios:
\begin{equation}\label{E:RrJ}
\dfrac{ Q(J_\epsilon u) }{ || J_\epsilon u ||^2_{L_2(M)} } \leq \big(1 + o(1) \big)  \dfrac{ Q_\epsilon( u ) }{ || u||^2_{L_2(M_\epsilon)}}
\end{equation}

\begin{equation}\label{E:RrK}
\dfrac{ Q_{\epsilon} ( K_\epsilon u) }{ || K_\epsilon u ||_{L_2(M_\epsilon)}^2} \leq \big( 1 + o(1) \big)  \dfrac{ Q(u) }{ || u ||_{L_2(M)}^2}
\end{equation}

Let now $W_n\subset \mathcal{G}^1$ and $W_n^\epsilon\subset H^1(M_\epsilon)$ be $n$, such that
\begin{equation}\label{E:minmax}
\lambda_n=\mathop{max}\limits_{x\in W_n\setminus \{0\}}\frac{Q(x,x)}{(x,x)},
\end{equation}
and
\begin{equation}\label{E:minmax}
\lambda^\epsilon_n=\mathop{max}\limits_{x\in W^\epsilon_n\setminus \{0\}}\frac{Q_\epsilon(x,x)}{(x,x)},
\end{equation}
Due to the min-max description and inequalities (\ref{E:RrJ}) and (\ref{E:RrK}), one gets

\begin{equation}
%\begin{split}
\lambda_n \leq
\sup_{u \in J_\epsilon(W_n^\epsilon)} \dfrac{ Q(J_\epsilon u) }{ || J_\epsilon u ||^2_{L_2(M)}}
 \leq \big( 1 + o(1) \big)  \lambda_n^\epsilon,
%\end{split}
\end{equation}
and
\begin{equation}
%\begin{split}
\lambda_n^\epsilon \leq
\sup_{u \in K_\epsilon(W_n)} \dfrac{ Q_\epsilon(K_\epsilon u) }{ || K_\epsilon u ||^2_{L_2(M_\epsilon)}}
\leq \big( 1 + o(1) \big)  \lambda_n
%\end{split}
\end{equation}
Thus, $ \lambda_n  - \lambda_n^\epsilon = o(1)$, which proves the theorem. $\Box$

The long technical task, to be addressed elsewhere, consists in proving the following statement:

\begin{theorem} \label{T:exist}
Let $M$ be an open book structure and its fattened partner $\{M_\epsilon\}_{\epsilon \in (0, \epsilon_0]}$ as defined before.
Let $A$ and $A_\epsilon$ be operators on $M$ and $M_\epsilon$ as in Definitions \ref{P:QeAe} and \ref{Ae}. There exist averaging operators $\{J_\epsilon\}_{\epsilon \in (0, \epsilon_0]}$ and extension operators $\{K_\epsilon\}_{\epsilon \in (0, \epsilon_0]}$ as stated in Definitions \ref{D:Je} and \ref{D:Ke}.
\end{theorem}

This leads to the main result of this text:

\begin{theorem} \label{T:main}
Let $M$ be an open book structure and its fattened partner $\{M_\epsilon\}_{\epsilon \in (0, \epsilon_0]}$.
Let $A$ and $A_\epsilon$ be operators on $M$ and $M_\epsilon$ as in Definitions \ref{P:QeAe} and \ref{Ae}.

Then, for any $n$ $$\lambda_n(A_\epsilon) \mathop{\rightarrow}_{\epsilon \to 0} \lambda_n(A).$$
\end{theorem}

%%%%%%%%%%%%%%%%%%%%%%%%%%%%%%%%%%
\section{Conclusions and final remarks}\label{S:remarks}
%%%%%%%%%%%%%%%%%%%%%%%%%%%%%%
\begin{itemize}
  \item As the quantum graph case teaches \cite{KZ2,Post}, allowing the volumes of the fattened bindings to shrink when $\epsilon\to 0$ slower than those of fattened pages, is expected to lead to interesting phase transitions in the limiting behavior. This is indeed the case, as it will be shown in yet another publication.
  \item It is more practical to allow presence of zero-dimensional strata (corners). The analysis and results get more complex, as we hope to show in yet another work, with more types of phase transitions.
  \item Resolvent convergence, rather than weaker local convergence of the spectra, as done in \cite{Post} in the graph case, would be desirable and probably achievable.
  \item One can allow some less restrictive geometries of the fattened domains.
  \item The case of Dirichlet Laplacian is expected to be significantly different in terms of results and much harder to study, as one can conclude from the graph case considerations \cite{Grieser}.
\end{itemize}
%%%%%%%%%%%%%%%%%%%%%%%%%%%%%%%%
\section{Acknowledgments}\label{acknow}
%%%%%%%%%%%%%%%%%%%%%%%%

The work of both authors was partially supported by the NSF DMS-1517938 Grant.


\begin{thebibliography}{99}


%%%%%%%%%%%%%%%%%%%%%%%%%%%%%%%
\bibitem{Arnold} Arnold, V. I. ;  Gusein-Zade, S. M. ;  Varchenko, A. N.  \emph{Singularities of differentiable maps. Volume 1. Classification of critical points, caustics and wave fronts}, Birkh\"{a}user/Springer, New York,  2012.

\bibitem{BerKuc} G. Berkolaiko and P. Kuchment, \emph{Introduction to Quantum Graphs}, AMS 2013.

\bibitem{DellAn15}G.~Dell'Antonio, A.~Michelangeli, Dynamics on a graph as the limit of the dynamics on a "fat graph''.  \emph{Mathematical technology of networks},
 49--64, Springer Proc. Math. Stat., 128, Springer, Cham,  2015.
		
\bibitem{DellAn07}G.~Dell'Antonio,  Dynamics on quantum graphs as constrained systems.
 \emph{Rep. Math. Phys.}  59  (2007),  no. 3, 267--279.
		
\bibitem{DellAn06}G.~Dell'Antonio,  L.~Tenuta, Quantum graphs as holonomic constraints.
\emph{ J. Math. Phys}.  47  (2006),  no. 7, 072102, 21 pp.

\bibitem{EE}  D. E. Edmunds and W. Evans, \textit{Spectral Theory and
Differential Operators}, Oxford Science Publ., Claredon Press, Oxford, 1990.

\bibitem{EH}  W. D. Evans and D. J. Harris, Fractals, trees and the Neumann
Laplacian, Math. Ann. 296(1993), 493-527.

\bibitem{AnGr} P. Exner, J. Keating, P.~Kuchment, T.~Sunada, and A.~Teplyaev (Ed.), \emph{Analysis on Graphs and its Applications}, Proc. Symp. Pure Math., AMS,2008.
%
\bibitem{EP} P. Exner and O. Post, Convergence of spectra of graph-like thin manifolds, J. Geom. Phys. \textbf{54} (2005), no. 1, 77-115.

\bibitem{ES}  P. Exner, P. Seba, Electrons in semiconductor microstructures:
a challenge to operator theorists, in \textit{Proceedings of the Workshop on
Schr\"{o}dinger Operators, Standard and Nonstandard} (Dubna 1988), World
Scientific, Singapore 1989; pp. 79-100.

\bibitem{FK1}  A. Figotin and P. Kuchment, Band-gap structure of the
spectrum of periodic and acoustic media. I. scalar model, SIAM J. Applied
Math. \textbf{56} (1996), no.1, 68-88.

\bibitem{FK3}  A. Figotin and P. Kuchment, Band-gap structure of the
spectrum of periodic and acoustic media. II. 2D photonic crystals, SIAM J.
Applied Math. \textbf{56} (1996), 1561-1620.

\bibitem{FK4}  A. Figotin and P. Kuchment, 2D photonic crystals with cubic
structure: asymptotic analysis, in \textit{Wave propagation in Complex Media}%
, G. Papanicolaou (Editor), IMA Volumes in Math. and Appl., \textbf{96} (1997),
23-30.

\bibitem{FK5}  A. Figotin and P. Kuchment, Spectral properties of classical
waves in high contrast periodic media, SIAM J. Appl. Math. \textbf{58} (1998), no.2,
683- 702.

\bibitem{Fr}  M. Freidlin, \textit{Markov processes and differential
equations: asymptotic problems}, Lectures in Mathematics ETH Z\"{u}rich,
Birkh\"{a}user Verlag, Basel, 1996.

\bibitem{FW}  M. Freidlin and A. Wentzell, Diffusion processes on graphs and
the averaging principle, Annals of Probability, \textbf{21} (1993), no.4, 2215-2245.

\bibitem{FWopb}  M. Freidlin and A. Wentzell, Diffusion processes on an open book and the
averaging principle, Stochastic Processes and their Applications \textbf{113} (2004), 101 -- 126.

\bibitem{FH} R. Froese and I. Herbst. Realizing holonomic constraints in classical and
quantum mechanics. Comm. Math. Phys. \textbf{220} (2001), 3, 489-535.

\bibitem{GeP}  N. Gerasimenko and B. Pavlov, Scattering problems on
non-compact graphs, Theor. Math. Phys., 75(1988), 230-240.

\bibitem{strat} M. Goresky, \emph{Stratified Morse Theory}, Springer Verlag 1988.

\bibitem{Grieser} D.~Grieser, Thin tubes in mathematical physics, global analysis and spectral geometry, in \cite{AnGr}, pp. 565--593, 2008.

\bibitem{KuPBG}  P. Kuchment, The Mathematics of Photonics Crystals, in \textit{%
Mathematical Modeling in Optical Science}, G. Bao, L. Cowsar, and W. Masters
(Editors), SIAM, 2001.

\bibitem{KuISAAC}  P. Kuchment, Differential and pseudo-differential operators on graphs as models of mesoscopic systems,  \emph{Analysis and applications—ISAAC 2001 (Berlin),
 7--30}, Int. Soc. Anal. Appl. Comput., Kluwer Acad. Publ., Dordrecht,  2003.
		
\bibitem{KZ2}  P. Kuchment, and H. Zeng, Asymptotics of spectra of Neumann Laplacians in thin domains.
 \emph{Advances in differential equations and mathematical physics (Birmingham, AL, 2002)},
 199--213, Contemp. Math., 327, Amer. Math. Soc., Providence, RI,  2003.
		
\bibitem{WRM}  P. Kuchment, Graph models for waves in thin structures,
 \emph{Waves Random Media } \textbf{12}  (2002),  no. 4, R1--R24.

\bibitem{KK1}  P. Kuchment and L. Kunyansky, Spectral properties of high
contrast band-gap materials and operators on graphs, Experimental
Mathematics, \textbf{8} (1999), no.1, 1-28.

\bibitem{KZ} P. Kuchment and H. Zeng, Convergence of spectra of mesoscopic systems collapsing onto a graph, J. Math. Anal. Appl. \textbf{258} (2001), 671-700.

\bibitem{Lu} Yung Chen Lu, \emph{Singularity theory and an introduction to catastrophe theory}, Universitext.
Springer-Verlag, New York-Berlin,  1980.

\bibitem{Ma} V. Maz'ja, \textit{Sobolev spaces}, Springer-Verlag, Berlin,
1985.

\bibitem{MaPo} V. Maz'ya and S. Poborchi, \textit{Differential functions on bad domains},  World Scientific, New Jersey, 1997.

%\bibitem{Mi}  S. G. Mikhlin, \textit{Mathematical physics, an advanced course}, North-Holland, Amsterdam, 1970.
%
%
%\bibitem{M78} S. G. Mikhlin, Equivalent norms in Sobolev spaces and norms of extension operators, \textit{Sibirsk. Mat. Zh.}, (1978), 1141-1153.
%
%\bibitem{M79} S. G. Mikhlin, On the minimal extension constant for functions of Sobolev classes, \textit{Zap. Nauchn. Sem. Leningrad. Otdel. Mat. Inst. Steklov. (LOMI)} \textbf{90} (1979), 150-185.


%\bibitem{N1}  S.Novikov, Schr\"{o}dinger operators on graphs and topology, Russian Math Surveys, 52(1997), no. 6, 177-178.

%\bibitem{N2}  S.Novikov, Discrete Schr\"{o}dinger operators and topology, to appear in Asian Math. J., 2(1999), no. 4, 841-853.

%\bibitem{N3}  S.Novikov, Schr\"{o}dinger operators on graphs and symplectic geometry, in \textit{The Arnoldfest: Proceedings of a Conference in Honour of V. I. Arnold for his Sixtieth Birthday}, E. Bierstone, B. Khesin, A. Khovanskii, and J. E. Marsden (Editors), AMS, 1999.

\bibitem{Post} O. Post, \textit{Spectral analysis on graph-like spaces}, Springer-Verlag, Berlin, 2012.

\bibitem{Ranic} A. Ranicki, \textit{High-dimensional knot
theory. Algebraic surgery in codimension 2}, Springer Verlag, Berlin, 1998.

\bibitem{ReSi} M. Reed and B. Simon, \textit{Methods of modern mathematical physics I: functional analysis}, Academic Press, San Diego, 1980.

\bibitem{RS1}  J. Rubinstein and M. Schatzman, Spectral and variational
problems on multiconnected strips, C. R. Acad. Sci. Paris Ser. I Math.
\textbf{325} (1997), no.4, 377-382.

\bibitem{RS2}  J. Rubinstein and M. Schatzman, Asymptotics for thin
superconducting rings, J. Math. Pures Appl. \textbf{77} (1998), no. 8, 801--820.

\bibitem{RS3}  J. Rubinstein and M. Schatzman, On multiply connected
mesoscopic superconducting structures, S\'{e}min. Th\'{e}or. Spectr.
G\'{e}om., no. 15, Univ. Grenoble I, Saint-Martin-d'H\`{e}res, 1998, 207-220.

\bibitem{RS4}  J. Rubinstein and M. Schatzman, Variational problems on
multiply connected thin strips I: basic estimates and convergence of the
Laplacian spectrum, Arch. Rational Mech. Anal. (2001), 160-271.

\bibitem{RueS}  K. Ruedenberg and C. W. Scherr, Free-electron network model
for conjugated systems. I. Theory, J. Chem. Physics, \textbf{21} (1953), no.9,
1565-1581.

\bibitem{S}  Y. Saito, The limiting equation of the Neumann Laplacians on
shrinking domains, Elec. J. Diff. Eq. \textbf{2000} (2000), No. 31, 1-25.

\bibitem{MS}  M. Schatzman, On the eigenvalues of the Laplace operator on a
thin set with Neumann boundary conditions, Applicable Anal. \textbf{61} (1996),
293-306.

		
\bibitem{Whit} H.~Whitney, \emph{Collected papers. Vol. II.}, Birkh\"{a}user Boston, Inc., Boston, MA,  1992.

\bibitem{Winkel} H. E. Winkelnkemper, Manifolds as open books, Bull. AMS \textbf{79} (1973), 45--51.



\end{thebibliography}
\end{document}